# RF-MEMS Switched Varactors for Medium Power Applications

F. Maury[1], A. Pothier[1], A. Crunteanu[1], F. Conseil[2] et P. Blondy[1]
[1] XLIM UMR 6172
123 avenue Albert Thomas, 87060 Limoges cedex
[2] MBDA France
20 Rue Grange Dame Rose BP150, 78141 Vélizy-Villacoublay cedex

*Abstract-* **In RF (Radio Frequency) domain, one of the limitations of using MEMS (Micro Electromechanical Systems) switching devices for medium power applications is RF power. Failure phenomena appear even for 500 mW. A design of MEMS switched capacitors with an enhanced topology is presented in this paper to prevent it. This kind of device and its promising performances will serve to fabricate a MEMS based phase shifter able to work under several watts.**

## I. Introduction

Many researches about MEMS trend to exploit their well-known and large advantages. Cost and size reduction enables MEMS sensors to be focus on. Some of them have demonstrated their mass production feasibility thanks to their CMOS compatibility. Nowadays, micro actuators are studied to follow the same way in telecommunication domain. Those which are driven by electrostatic actuation are the most investigated. This kind of structures has an easy way to build; a simple difference of potential between two metallic layers is enough to create an attraction force for actuation. The most mature of them have been already commercialized, only tens of volts enable them to switch [1]. Their interesting performances promise to be very suitable for a large kind of control circuits.

For example, MEMS switched capacitors are used for their low series resistances and low-drive power requirements. For wireless communication and particularly radar applications, switched-line phase shifters using MEMS fixed-fixed beams have demonstrated good degree/ loss ratios. Nevertheless, the choice of these MEMS structures is compromised for high and even medium power applications. For microwave signals beyond 1 Watt, most of MEMS switched capacitors fail in hot switching operation. In this operating way, the RF signal is applied continually and, contrary to cold switching, it is not interrupted when the MEMS beams are activated. This is a better manner of working for users who don't have to care about RF signal source control.

The most critical power failure phenomenon appears by this way is called RF latching. In down state position, the electrostatic pressure generated by RF power prevents the beam to recover the up state when the DC actuation signal is stopped. To avoid this event, the mobile structure must have a better stiffness to lead to a restoring force capable of frustrate this electrostatic force [2].

## II. Device design

In the research institute XLIM in Limoges, we have proposed a design of MEMS switched capacitors with an enhanced topology offering a better opposition to RF power forces compared to traditional architecture [3]. This geometry has been especially studied to consider electrostatic pressures generated by high RF power signal. The actuation part of cantilevers and their contact area are well separated and have been designed to be almost fully independent. That's why, the architecture of the beam consists in two different areas: a large actuation part and a smaller contact tip as shown on Fig.1.

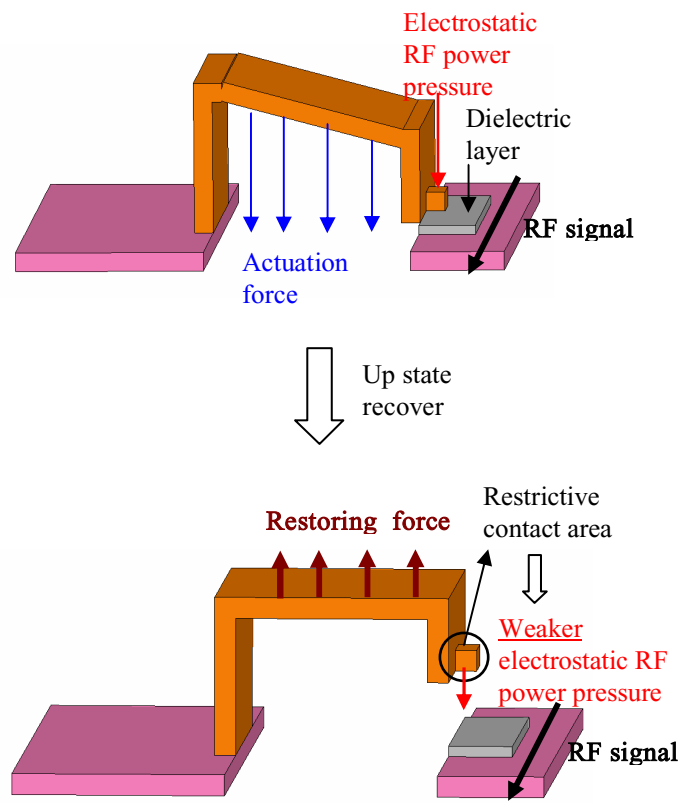

Fig. 1. Principle of proposed cantilever based switched varactor.

ISBN: 978-2-35500-006-5

Hence, when a RF signal is applied across the component, due to its electric field, a static attractive electrostatic pressure is induced on the movable beam. In fact, this pressure mainly depends both on the RF signal magnitude, on the localized area geometry of the beam where this field is concentrated (the beam tip in our case) and also on the distance separating this area from the RF line where the signal propagates.

To recover up state position, when the actuation command is interrupted, the beam must have a restoring force greater than the electrostatic RF power attraction force. Designing a restricted beam tip area allows to reduce the effect of electrostatic pressure induced by the RF signal which is really critical in down state position when the tip is put into contact with RF line. Indeed, contrary to the up state position, the induced RF electric field is much higher in down state since the air gap is almost reduced to zero, only a thin dielectric layer isolates the beam tip from the RF line. That is why hot switching operation is the operating mode where power failure first appears. Thanks to this specific tip geometry, the power handling of the component could be highly improved up to several Watts on 50 Ohms impedance.

The other particularity of the proposed beam geometry is the difference of gaps between the actuation area and the contact tip. Hence in the beam down state position, a higher air gap on the actuation area allows improving the contact force on the tip and avoiding in the same way some collapse phenomenon between the beam and its pull-down electrode after the tip has made contact on the RF line. Thus, the component reliability could be grandly improved since dielectric charging troubles generally observed in contact mode electrostatic actuators is strongly reduced.

III. TEST CELLS

Several switched capacitors geometries have been designed and analytical calculus have predicted that most of them should be able to handle RF signal power of 4 Watts at least. To evaluate the real power handling capabilities under hot switching conditions, fabricated devices have been tested for different level of RF power monitoring their minimum DC release voltages before recovering the cantilever up state. An example of cell which have been tested is shown Fig.2 and Fig.3 presents there is no significant voltage change up to 18 Watts under 50 Ohms revealing no failure phenomenon occurs at least until this RF power level.

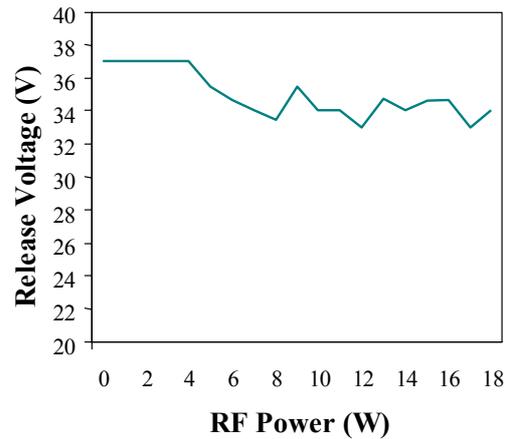

Fig. 3. Measures of DC actuation release voltages.

To our knowledge, there is no MEMS based phase shifter in literature able to operate under several Watts in hot switching conditions. The presented switched capacitor concept could be implemented in a DMTL (Distributed MEMS Transmission Line) to reach such power handling. To design such phase shifter a high impedance microwave transmission line is generally used periodically loaded with tens or more of varactors, the principle of a DMTL phase shifter topology is shown Fig.4 and a fabricated testing cell is presented Fig.5.

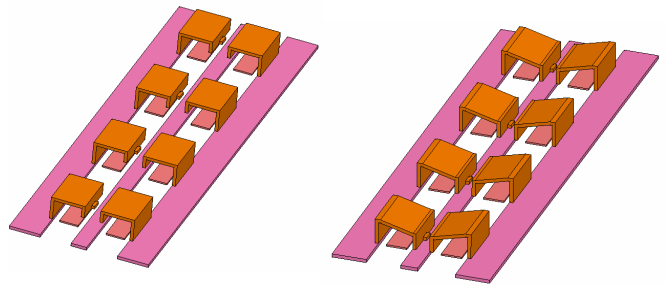

Fig. 4. Principle of a DMTL phase shifter topology.

The phase shift is achieved by turning on several of the switched capacitors distributed along the line allowing a decrease of the line impedance and so meaningful change on the RF signal velocity. Actually, the available phase shift increases with the number of cantilevers placed along the line.

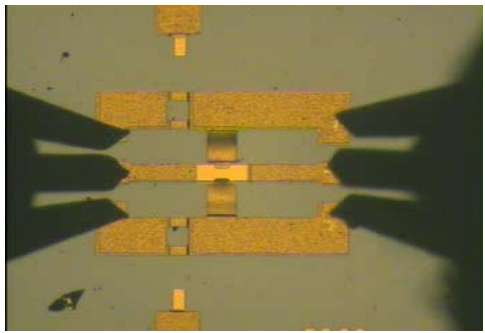

Fig. 2. Testing cell.

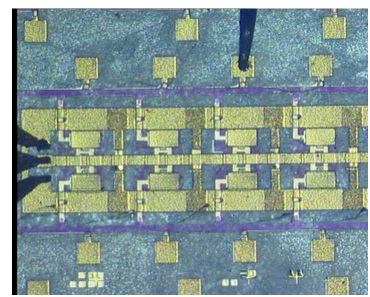

Fig. 5. Picture of a fabricated phase shifting cell.

Thus several switched capacitors can be actuated in the same time as a single element (a bit) in order to reach well defined phase shift value: 0°, 90°, 180°, 270° as example on a 2 bit phase shifter. Compared to analog phase shifters, this approach guarantees a very low sensitivity to phase noise and better performance reproducibility.

## IV. CONCLUSION

For instance, several elementary phase shifting cells (1-bit topology) have been fabricated. Each cell is based on a coplanar waveguide loaded with 2 cantilevers symmetrically placed at each side. The RF performances and the power handling of these structures are currently under test. The last results will be presented during the conference.